\documentclass{aa}

\usepackage{txfonts}
\usepackage{epsfig} 
\usepackage{amssymb}
\usepackage{graphics}
\usepackage{natbib}
\bibpunct{(}{)}{;}{a}{}{,} 

\newcommand{\fxo}{$f_{\rm X}/f_{\rm opt}$}
\newcommand{\fx}{$f_{\rm X}$ }
\newcommand{\cgs}{erg~cm$^{-2}$~s$^{-1}$}

\newcommand{\xmm}{2XMMp}
\newcommand{\xmmc}{2XMMp \rm catalogue}
\newcommand{\gf}{{\it Galaxy Fraction}}
\newcommand{\src}{2XMM J123204+215255}
 
\begin{document}
\title{An extreme EXO: a type 2 QSO at $z=1.87$}
\author{A.~Del Moro
\inst{1}
\and M.~G.~Watson
\inst{1}
\and S.~Mateos
\inst{1}
\and M.~Akiyama
\inst{2}
\and Y.~Hashimoto
\inst{3}
\and N.~Tamura
\inst{4}
\and K.~Ohta
\inst{5} 
\and F.~J.~Carrera
\inst{6}
\and G.~Stewart
\inst{1}}

\institute{XROA-University of Leicester, University Road, Leicester LE1 7RH, UK
\and Astronomical Institute, Tohoku University, Sendai 980-8578, Japan
\and South African Astronomical Observatory, Observatory Road, Cape Town 7539, South Africa
\and Subaru Telescope, National Astronomical Observatory of Japan, Hilo, HI 96720
\and Department of Astronomy, Kyoto University, Kyoto 606-8502, Japan 
\and Instituto de F\'\i sica de Cantabria (CSIC-UC), Avenida de los Castros, 39005 Santander, Spain}

\abstract{}{We aim to understand the multi-wavelength properties of \src, the source with the most extreme
X-ray-to-optical flux ratio amongst a sample of bright X-ray selected EXOs drawn from a cross-correlation of the \xmmc\ with 
the SDSS-DR5 catalogue.}
{We use \xmm\  X-ray data, SDSS-DR5, NOT and UKIRT optical/NIR photometric data and Subaru MOIRCS IR spectroscopy to study the 
properties of \src. We created a model SED including an obscured QSO and the host galaxy component to constrain the optical/IR 
extinction and  the relative contribution of the AGN and the galaxy to the total emission.}
{\src\ is a bright X-ray source with f$_{\rm X}\approx10^{-12}$ \cgs\ (2-10 keV energy band) which has no detection down to a 
magnitude $i'>25.2$. NIR imaging reveals a faint $K-$band counterpart and NIR spectroscopy shows a single broad 
(FWHM $\simeq5300$\ km/s) emission line, which is almost certainly H$\alpha$ at $z=1.87$. The X-ray spectrum shows evidence of 
significant absorption ($N_H>10^{23}\ \rm cm^{-2}$), typical of type 2 AGN, but the broad H$\alpha$ emission suggests a type 1 
AGN classification. The very red optical/NIR colours ($i'-K>5.3$) strongly suggest significant reddening however. We find that 
simple modelling can successfully reproduce the NIR continuum and strongly constrain the intrinsic nuclear optical/IR extinction
to A$_V\approx 4$, which turns out to be much smaller than the expected from the X-ray absorption (assuming Galactic gas-to-dust
ratio).}{}
\label{abs}

\maketitle

\section{Introduction}
\label{intro}

Over the past few years, extensive studies have been carried out to determine the complete census of the active 
galactic nuclei (AGN) population, in particular the bright, obscured objects (optically type 2 QSOs), invoked by the 
synthesis models of the X-ray background (XRB, \citealt{setti1989,comastri1995,comastri2001,gilli2001,gilli2007}).
The deepest surveys carried out to date by {\it Chandra} and XMM-Newton have resolved $>90\%$ \citep{mushotz2000,
hasinger2001, alexander2003} of the XRB spectrum at energies below 5 keV, however  a large number of the obscured AGN 
are still missing. For these reasons, many studies have been addressed to reveal this elusive class of objects, primarily 
focused on the X-ray band, but also involving multi-wavelength studies \citep{zakamska2003,zakamska2004,alejo2006}.

In previous studies, the X-ray-to-optical-flux ratio (\fxo\footnote{\fxo\ is defined here as  $\log(f_{\rm
X}/f_{\rm opt})=\log f_{\rm 2-10~keV}+r'/2.5+5.5$ \citep[e.g.][]{civano2005}, where $f_{\rm 2-10~keV}$ is the X-ray flux (\cgs)
and $r'$ is the Sloan magnitude in the $r'$ band ($AB$).}), defined here as the ratio between the X-ray flux (usually  in the 
2--10 keV energy band) and $R-$band flux, has been commonly used to provide a first classification of the X-ray  population
\citep{maccacaro1988}: normal galaxies have typically \fxo\ $\lesssim0.1$ \citep[e.g.][]{giacconi2001,lehmann2001} while the 
dominant X-ray selected AGN population has $0.1<$ \fxo\ $<10$ \citep[e.g.][]{akiyama2000,alexander2001,fiore2003}; obscured AGN 
typically have \fxo\ $\sim10$ or above \citep[e.g.][]{mignoli2004}. 

Recently, a new interesting class of objects with extreme values of the X-ray-to-optical flux ratio (\fxo\ $>10$) has 
been found from deep X-ray surveys; these objects, called {\it Extreme X-ray-to-Optical ratio sources} 
\citep[EXOs,][]{cocomero2004}, are usually detected in X-rays and in near-IR, but completely undetected in the optical 
bands. Although a few turn out to be BL Lac objects, high-redshift clusters of galaxies or rare Galactic objects (e.g.
X-ray binaries, cataclysmic variables, isolated neutron stars or ultraluminous X-ray sources), the majority are
almost certainly type 2 AGN \citep{fiore2003}, where the nucleus is heavily obscured by dust in the UV/optical/IR
whilst the X-ray flux, although strongly absorbed at low energies (below $\sim 2$ keV), is much less affected by 
absorption above $\sim 2$ keV.

As expected in this scenario, a significant fraction of the EXO population shows the extremely red colours of EROs 
({\it extremely red objects}, typical $R-K\ge5$, \citealt{elston1988}; see \citealt{fiore2003,mignoli2004}). EXOs with 
optical-NIR colours of EROs are thus amongst the best candidates to be highly absorbed, highly obscured AGN, i.e. type 2 AGN.  

However, almost by definition, sources with such extreme \fxo\ have faint optical magnitudes and this is even more true 
for the faint X-ray sources found in the deep, pencil-beam surveys \citep{cocomero2004,civano2005}. At such faint X-ray
fluxes (\fx$=10^{-15}-10^{-16}$ \cgs), sources with \fxo\ $>10$ have R magnitudes $\sim 25.5-28$, so the optical follow-up 
observations are very difficult or impossible. Studies performed at higher X-ray fluxes (\fx$=10^{-14}-10^{-13}$ \cgs,
e.g.\ \citealt{fiore2003,mignoli2004,severgnini2006}), for which optical data are available, have classified more than a 
half of their EXO sources as type 2 QSOs.

In this paper, we report the results obtained from the multi-wavelength analysis of 2XMM J123204+215255, the most 
extreme source amongst a sample of 130 bright (\fx$\ge10^{-13}$ \cgs) X-ray selected EXOs (Del Moro et al., in 
preparation) from the \xmmc\ (the pre-release of the Second XMM-Newton Serendipitous Source Catalogue, \citealt{watson2008}). 
The whole sample is described in \S\ \ref{sam}. The X-ray and optical/NIR properties of \src\ are discussed in \S\ 
\ref{xray} and \S\ \ref{opt}, respectively; a detailed description of the NIR spectrum is presented in  \S\ \ref{ir}. An SED 
model analysis and results are discussed in \S\ \ref{discus}, followed by the conclusions (\S\ \ref{conc}). Throughout the 
paper we assume a cosmological model with $H_0=70\ \rm km\ s^{-1}\ Mpc^{-1}$, $\Omega_M=0.27$ and $\Omega_{\Lambda}=0.73$ 
\citep{spergel2003}.

\section{Observations and results}
\subsection{The EXO sample}
\label{sam}
The object we present in this paper is part of a sample selected from a cross-correlation between the \xmm\footnote{Although
this object was originally selected from the \xmm, we adopt the nomenclature and catalogue parameters from the 2XMM 
catalogue throughout as 2XMM effectively supersedes 2XMMp.} and the {\it Sloan Digital Sky Survey, Data Release 5} 
(SDSS-DR5, \citealt{adelmanmc2007}) catalogues, which yields about 20000 secure matches. SDSS optical counterparts have been 
selected on the basis of positional matching, using an adaptation of the {\it likelihood ratio} method (e.g. \citealt{ss1992})
to optimise the choice of correct counterpart and minimise the contamination by chance matches. Full details of our approach 
will appear in Del Moro et al., in preparation.

Our sample consists of relatively bright X-ray sources with \fx$\ge10^{-13}$ \cgs\ (0.2--12 keV energy band) and extreme 
X-ray-to-optical flux ratio (\fxo\ $>10$, in the 2--10 keV band). Each XMM and SDSS image of the sample objects has been 
carefully checked by eye in order to reject all the problematic cases: spurious X-ray detections, sources close to the edge of 
the XMM-Newton field of view, extended X-ray sources\footnote{As we are looking for AGN, we want to select only X-ray 
point-like sources in our sample, so extended sources are considered as ``problematic cases" amongst our selection criteria.},
optical counterparts in big nearby galaxies (HII region emissions). Because we are selecting extreme X-ray-to-optical flux 
ratio objects, the counterparts of our sources are typically faint in the optical, at our sensitivity limits, 
making it difficult in some cases to differentiate between the possibility that the true match is fainter than the SDSS limit or
that the correct counterpart is a faint object with low likelihood ratio (and hence with a higher probability to be a
chance match). The effects of this ambiguity are that the \fxo\ values, in these cases, will be {\it underestimated} (i.e. if 
the counterpart is fainter than the SDSS limit). In this sense our sample is robust.

The sample resulting from these selection processes consists of 130 sources for which $\sim$30\% (including \src) have no 
optical counterpart in the SDSS imaging data within 7\arcsec\ from the X-ray position (see \S~\ref{xray}), down to a 
magnitude limit of $r'=22.5$.

\subsection{X-ray properties of \src}
\label{xray}
The source \src\ is the most extreme X-ray-to-optical flux ratio object amongst our EXO sample. It is a bright X-ray 
source with \fx$=8.8^{+0.7}_{-2.6}\times10^{-13}$ \cgs\ (in the 2--10 keV energy band). On the basis of the lack of an SDSS 
counterpart it has \fxo\ $>278$, already one of the highest ratios recorded, whilst as we show below, new data demonstrate the 
ratio to be \fxo\ $>3300$, making it the highest known X-ray-to-optical flux ratio source outside the Galaxy. The source 
has been detected in an XMM-Newton observation with target ``NGP Rift 3'' which was made in July 2001 with an exposure 
time of $\sim15$ ks with pn and $\sim20$ ks with MOS1 and MOS2. 
\begin{figure}
\centerline{
\includegraphics[scale=0.36,angle=0]{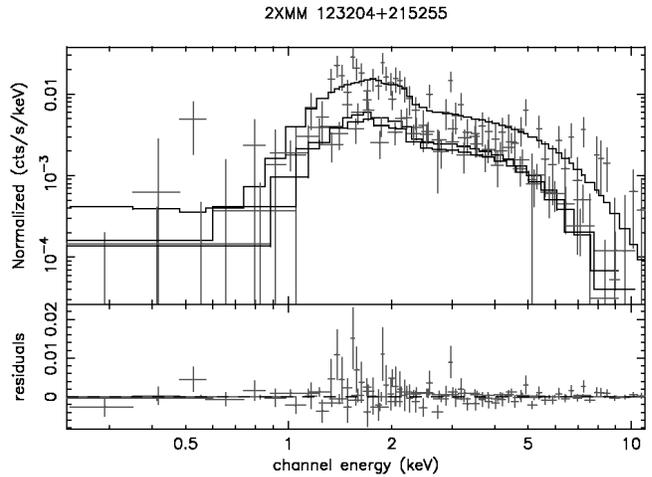}}
\caption{EPIC pn, MOS1 and MOS2 X-ray spectrum of 2XMM J123204+215255 (top panel): the model is represented by the black
lines; residuals are also shown (lower panel). The data are well represented by an absorbed power-law with a photon index 
$\Gamma=1.5\pm0.1$ and a hydrogen column density $N_H=1.8\times10^{23}\ \rm cm^{-2}$ at $z=1.87$.}
\label{fig.xspec}
\end{figure}

The statistical position error for \src\ is 0.3\arcsec. Taking into account the expected systematic error component for the 
2XMMp catalogue of 0.35\arcsec, determined from a comparison of catalogue positions with the SDSS-DR5 Quasar Catalog 
\citep[see][]{watson2008}, we expect the counterpart to lie within $\approx 1.4$\arcsec\ (99\% confidence). The 
comparison with the SDSS Quasar Catalog shows that there are a very small number of outliers at larger separations in excess 
of what is expected statistically (presumably indicating a larger systematic error component in these rare cases), but in no 
case the separation is $>7$ arcsec. We adopt this value as the most conservative upper limit on the possible separation of 
the true counterpart.

To study the X-ray spectrum of \src, the data have been processed using the standard SAS v.7.1.0 tasks (XMM-Newton Science 
Analysis System, \citealt{gabriel2004}). We extracted the spectrum of the source using an elliptical region to reproduce the 
shape of the Point Spread Function (PSF) of the XMM Telescopes at the source position. The corresponding background spectrum 
has been extracted using an annular region with radii 30\arcsec\ and 90\arcsec, centred on the source position, removing any 
other detected nearby sources. The data have been then filtered for high background intervals. The total EPIC number of 
counts in the 0.2--12 keV energy band after the filtering is 2234. The spectral analysis has been performed with XSPEC 
v.11.3.2 \citep{arnaud1996}, grouping the number of counts to a minimum of 10 counts per bin, in order to use the $\chi^2$ 
statistic. We find an acceptable fit to a simple model composed of a power-law plus Galactic and intrinsic absorptions; 
fixing $N_H^{gal}=2\times10^{20}\ \rm cm^{-2}$, the best fit parameters are photon index $\Gamma=1.7\pm0.2$ and hydrogen 
column density $N_H=(1.2\pm0.2)\times 10^{22}\ \rm cm^{-2}$ (assuming $z=0$), with $\chi^2/\rm d.o.f.=118.0/106$. 
\begin{figure*}[!t]
\centerline{
\includegraphics[scale=0.4]{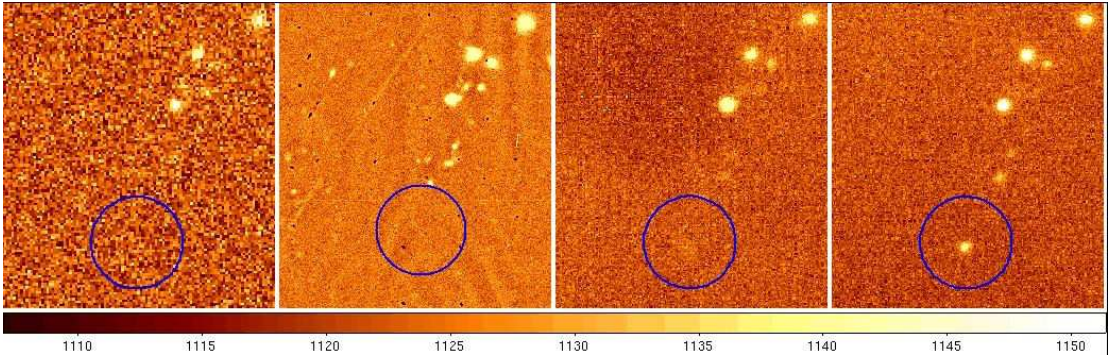}}
\caption{Optical/IR finding charts for 2XMM J123204+215255. From left to right: SDSS $r'$-band image, NOT 
$i'$-band stacked image, UKIRT-WFCAM $J$ and $K$ bands.The $K-$band image shows an apparently stellar counterpart ($K=19.9,
AB$). The circles have a radius of 7$\arcsec$ and are centred on the X-ray position.}
\label{fig.chart}
\end{figure*}
\begin{table*}
\caption{Multi-wavelength properties of \src.}
\label{table:1}
\begin{tiny}
\begin{tabular}{c c c c c c c c c c}
\hline\hline
\rule[-1.5mm]{0pt}{4ex}IAU Name& RA DEC& EPIC cts$^{a}$ & $f_{\rm 2-10~keV}$ & $L_{\rm 2-10~keV}$ &\fxo& $r'^{b}$ & $i'^{c}$ &
$J^{d}$ & $K^{d}$  \\
\rule[-1.5mm]{0pt}{3ex} &[h:m:s][d:m:s]& &$[10^{-13}$ \cgs] & [$10^{46}$ erg/s] & &[mag] & [mag]& [mag]&[mag]\\
\hline
\rule[-1.7mm]{0pt}{4ex}
2XMM J123204+215255 & 12:32:04.91+21:52:55.4 & 2234 & $8.8^{+0.7}_{-2.6}$ & 1.6 & $>$278 ($>$3300)$^*$& $>$22.5& $>$25.2 &
$>$22.2 & 19.9 \\
\hline
\end{tabular}

$^a$ Total pn+MOS1+MOS2 counts.\\
$^b$ SDSS ($AB$).\\
$^c$ NOT ($AB$).\\
$^d$ UKIRT-WFCAM ($AB$).\\
$^*$ Expected value derived from i-band limit, see text.
\end{tiny}
\end{table*}

Repeating the X-ray spectral analysis for $z=1.87$, anticipating the result presented below (\S~\ref{ir}), we obtained a best 
fit to the data with a photon index $\Gamma=1.5\pm0.1$ and a column density $N_H=(1.8\pm0.3)\times10^{23}\ \rm cm^{-2}$ 
(Fig.~\ref{fig.xspec}). We note the presence of marginally significant structure in the spectrum residuals at energy $\sim 
1.5-3$ keV which might be related to red-shifted line or edge features. The reality of these features clearly require 
confirmation at higher signal-to-noise.

\subsection{Optical/IR properties}
\label{opt}
As noted above (\S~\ref{sam}), there is no counterpart for \src\ in SDSS DR5 within 7\arcsec\ (\S~\ref{xray}) from the X-ray 
position in any of the SDSS bands ($u', g', r', i', z'$, \citealt{fukugita1996}), indeed the nearest SDSS object lies 
$\sim 30$\arcsec\ away.

We obtained deeper optical imaging of the field on 2007 February 9 on the 2.5m NOT ({\it Nordic Optical Telescope}). 
Four separate $i'-$band images were taken with the NOT ALFOSC camera with integration times of 1800 seconds each. 
Conditions were photometric with $\sim1$ arcsec seeing, but unfortunately the images are affected by bad fringing. As 
there were no sky flats available to correct properly for the fringing, an empirical approach was adopted in which a 
smoothed version of each image (with bright objects removed) is used to estimate the effective sky background plus 
fringing effects. The resultant background-subtracted images were then stacked to make the final image (Fig. 
\ref{fig.chart}). This technique does not fully remove the fringing effects, but reduces them to a modest level, 
significantly improving the sensitivity for faint objects. 

We find no detection of any counterpart within 7 arcsec of the position of \src\ with an estimated 5$\sigma$ limit $i'>25.2$ 
mag., based on the amplitude of the residual background fluctuations. The photometric calibration was established by comparing 
the count rates of a sample of objects detected in the NOT image with the SDSS-DR5 photometric catalogue. 

As the object is likely to be heavily obscured in the optical, the obvious next step is IR observations, where the effective 
dust obscuration will be lower. Follow-up $J-$band and $K-$band observations of \src\ have been obtained with the {\it UK 
Infrared Telescope--Wide Field Infrared Camera} (UKIRT--WFCAM) the $12^{th}$ of February 2007 with an exposure time of 1200 s 
in the $J-$band and $2\times1200$ s in the $K-$band. The IR images (Fig. \ref{fig.chart}) show an apparently stellar 
counterpart in the $K-$band with a magnitude $K=18.0\pm 0.05$ (Vega, $K=19.9\pm 0.05\ {\rm in}\ AB$ magnitude system, 
\citealt{okegunn1983}; see \citealt{hewett2006}) and a very marginal detection in the $J-$band ($J\gtrsim21.3$, Vega). All the 
values are reported in table \ref{table:1}. The IR counterpart lies $\approx 1$\arcsec\ from the X-ray centroid, consistent with
the position errors discussed in \S~\ref{xray}. The IR detection allows us to place limits on the optical/IR colours of the 
source: $r'-K>2.6$ and $i'-K>5.3$ ($AB$). As the optical emission is likely to be dominated by a high redshift galaxy, as 
discussed below, it is likely that $r'-i'>0$, which means that the $r'$ magnitude should be of the order of $r'\gtrsim25.2$. If 
that is true, the $r'-K$ colour becomes greater than 5.3 ($AB$), corresponding to $r'-K>7$ in the Vega system, which is 
extremely red even for EROs (\S\ \ref{intro}). 

\subsection{IR spectrum}
\label{ir}
\src\ was observed in the $H+K$ bands with the {\it Multi-Object Infrared Camera and Spectrograph} (MOIRCS,
\citealt{ichikawa2006}) at the Subaru Telescope in March 2007. The observation was performed in long-slit spectroscopy mode 
using the $HK500$ grism and 0.8\arcsec\ slit width, under a seeing of 0.6\arcsec. The spectrum covers a wavelength range of 
13000--23000 \AA\ with a spectral resolution of $\approx40$ \AA\ ($\sim$ 5 pix) FWHM estimated from the width of OH 
airglow lines.

\begin{figure}
\centerline{
\includegraphics[scale=0.33, angle=-90]{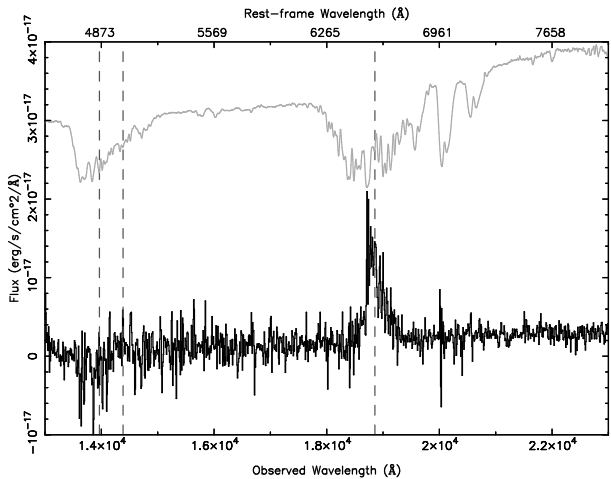}}
\caption{The Subaru MOIRCS IR spectrum of \src\ (black) showing a single broad emission line ($\rm FWHM=5280\pm331\ \rm km/s$), 
likely to be H$\alpha$ at $z=1.87$. The grey spectrum shows the ratio between raw and final flux-calibrated spectrum. 
The ratio is dominated by the effects of atmospheric absorption and thus provides an indication of the main regions where
such features are significant. The dashed vertical lines indicate the central wavelengths of H$\alpha$ (see \S~\ref{ir}) and the
expected H$\beta$ and [O{\sc III}] emission lines for this redshift. On the top axis the rest-frame wavelengths are plotted.}
\label{fig.irspec}
\end{figure}

We took 6 target frames with 7 minutes exposure each, thus the total on-source integration time was 42 minutes. The 
telescope was dithered after every single exposure and the target was observed at different positions along the slit
($1.5\arcsec$ or $3\arcsec$ from the slit centre). After sky subtraction, carried out by subtracting 
adjacent object frames from each other, flat-fielding was performed using dome-flat frames. These sky-subtracted and 
flat-fielded frames were then shifted, mirrored and combined to determine the average intensity value at each pixel with a 
3$\sigma$ clipping algorithm. Atmospheric absorption correction and flux calibration have been performed using the spectrum of 
a bright star (HD 119496, spectral type of A2V). The standard star spectrum was taken at the end of the night at a similar 
airmass to the target and with the same instrumental setup.

The observed spectrum (Fig. \ref{fig.irspec}) reveals a red continuum emission, which is well represented by 
$f_{\lambda} \propto \lambda$. Using this form and extrapolating it out to the wavelengths longer than 23000 \AA, the
$K-$band magnitude of this source is estimated to be $19.7 \pm 0.1$ mag in the $AB$ system without any corrections (the emission 
line is at the edge of the $K-$band and its effect on this estimate can be ignored). This is consistent with the UKIRT-WFCAM 
photometry ($K=19.9 \pm 0.05$, see \S\ \ref{opt} and Table \ref{table:1}) within the uncertainties. An estimated $\sim$10\% of 
the total flux from the object falls outside the slit width in the $0.6\arcsec$ seeing conditions of the observation. As the
slit losses are uncertain and the magnitude estimates from the spectroscopy are consistent with the UKIRT imaging, we therefore 
do not apply any slit-loss corrections to the emission line fluxes estimated below, but note that they have $\sim$10-20\%
systematic uncertainty.

As shown in Fig. \ref{fig.irspec}, a single broad emission line (FWHM$=5280\pm331$ km s$^{-1}$) is clearly visible 
on the red continuum in the observed spectrum at $18837\pm7$ \AA\ (line centroid). The line flux is $f=(5\pm0.2)\times10^{-16}$ 
\cgs\ (uncorrected for absorption; Table \ref{table2}). The emission line centre and line width have been estimated by fitting 
a Gaussian to the spectrum in the vicinity of the line after the continuum was subtracted, whilst the line flux was estimated 
by directly integrating the continuum-subtracted line profile. The derived \emph{Equivalent Width} (EW) is EW $=2500\pm140$ \AA. 
All the uncertainties are estimated from the 1$\sigma$ errors in the Gaussian fitting process.

Although there are no other significant emission or absorption lines visible in the spectrum, the most likely 
interpretation is that the line detected is H$\alpha$. We also considered other possible interpretations for this line:
\begin{itemize}
\item[i)] if it were one of the Paschen series the implied redshift would be $z<1$ which seems unlikely as the host 
galaxy is not detected in the optical (and would thus be a very low luminosity object) and the detected line flux would be much 
higher than expected; 
\item[ii)] if it were MgII the redshift would be $z\sim5.7$ which would make the X-ray absorption 
$N_H\approx1.5\times10^{24}\ \rm cm^{-2}$ (Compton thick). However, no strong Fe K$\alpha$ line is detected in the X-ray 
spectrum and the derived luminosity at this redshift is extremely high ($L_{\rm X}\approx1.8\times10^{47}\ \rm erg~s^{-1}$),
so we believe also this hypothesis is unlikely.
\end{itemize}
Assuming the H$\alpha$ identification is correct, the estimated redshift for the source is $z=1.87$. Even though the emission
line is just at the wavelength range where the spectrum is seriously affected by the atmospheric absorption lines between $H$ 
and $K$ bands, this can not explain the total absence of the [O{\sc III}] (expected from the Narrow Line Region) 
and the H$\beta$ emission lines. For those lines we estimate flux upper limits from our MOIRCS spectrum: 
$f_{\rm [O{\sc III}]}<3 \times 10^{-17}$ \cgs\ (assuming an FWHM $\le$ 500 km s$^{-1}$) and $f_{{\rm H}\beta}<1 \times10^{-16}$
\cgs\ (assuming the same FWHM as H$\alpha$). These limits are conservative values estimated at several times higher than the
formal 3$\sigma$ level (derived from empirical estimates of the noise in the vicinity of these lines) to allow for the 
systematic deviations from the expected smooth continuum. The estimated line ratio H$\alpha$/H$\beta$ is $>5$, which is
higher than the typical intrinsic unreddened value for AGN (H$\alpha$/H$\beta\approx3.5$, e.g. \citealt{ward1988}). Noting that 
the H$\beta$ flux is a conservative upper limit, this already indicates significant extinction in this object. We also note the
clear asymmetry in the H$\alpha$ line profile for which we have no obvious explanation, although an incomplete correction for 
the severe atmospheric absorption may be a possible reason. A narrow  [N {\sc II}] $\lambda6583$ line may also be present in the
spectrum and contribute to the asymmetry of the H$\alpha$ profile. The presence of this line, assuming it has FWHM $\le$ 500 km
s$^{-1}$, cannot however explain all the red wing, given the H$\alpha$ line width.


\section{Discussion}
\label{discus}
For a redshift $z=1.87$, the intrinsic X-ray luminosity for the source is $L_{\rm X}=1.6\times10^{46}$~erg~s$^{-1}$ 
(2--10 keV rest-frame) and the X-ray spectrum has a large column density ($N_H\approx2\times10^{23}\ \rm 
cm^{-2}$; \S~\ref{xray}). The X-ray properties of \src\ are thus those of a very luminous, heavily absorbed AGN, with 
spectral parameters typical for a type 2 QSO. 

In contrast the IR spectrum shows a broad H$\alpha$ emission line and the IR counterpart detected in the $K-$band has a 
probable stellar morphology, characteristics of a type 1 object. To further complicate the story the predicted emission 
line fluxes are much higher than those observed (see Table \ref{table2}). Using the correlation between hard X-ray 
luminosities and H$\alpha$ and [O{\sc III}] luminosities from \citet{panessa2006}:  
\begin{eqnarray}
\log{L_{\rm X}}&=&(1.06\pm0.04)\log{L_{\rm H\alpha}}+(-1.14\pm1.78) \nonumber  \\
\log{L_{\rm X}}&=&(1.22\pm0.06)\log{L_{\rm [O{\sc III}]}}+(-7.34\pm2.53) \nonumber
\end{eqnarray}
the predicted flux for H$\alpha$ is $f_{\rm H\alpha}\sim1.7\times10^{-14}$ \cgs, which is about 30 times higher than
the detected line flux (uncorrected for absorption) and the predicted fluxes for [O{\sc III}] and H$\beta$ are 
$f_{\rm [O{\sc III}]}\sim2.9\times10^{-15}$ \cgs\ and $f_{\rm H\beta}\sim4.8\times10^{-15}$ \cgs\ (assuming the standard 
$\rm{H}\alpha/\rm{H}\beta$ ratio, e.g. \citealt{ward1988}), whereas these lines are not detected at all.
\begin{figure}[t!]
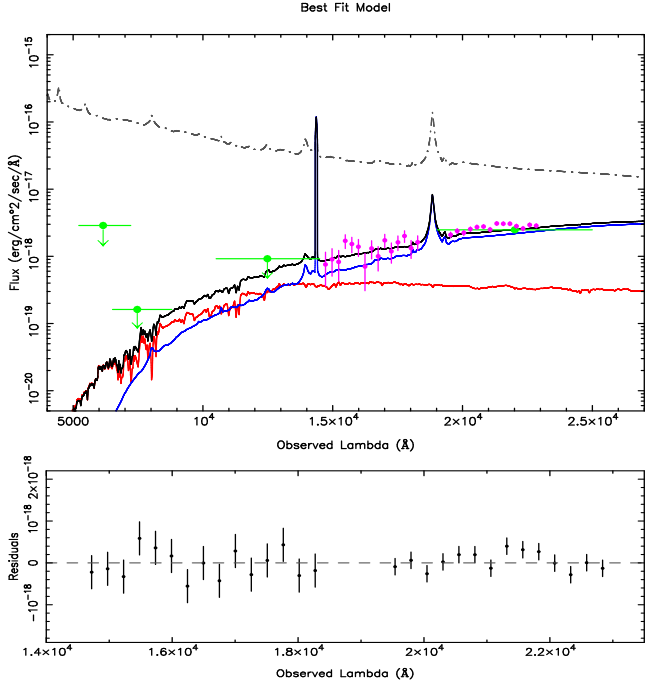

\centerline{
\includegraphics[height= 8.5 cm, width=6 cm,angle=-90]{9665fig4.ps}}
\centerline{
\includegraphics[height= 8.5 cm, width=3 cm,angle=-90]{9665fig5.ps}}
\caption{Best-fit model of the NIR spectrum (magenta points); green circles: photometric measurements and upperlimits 
of the $r'$, $i'$, $J$ and $K-$band magnitudes; red line: galaxy SED; blue line: absorbed QSO SED; grey dash-dotted line: 
unabsorbed QSO SED; black line: resulting SED (QSO + host galaxy). In the lower panel the residuals of the fitting are shown.}
\label{fig.modelfit}
\end{figure}

In order to see whether it was possible to reconcile the X-ray, optical and near IR characteristics of this object, we carried 
out simple modelling of the IR spectral continuum using a composite Spectral Energy Distribution (SED) with two 
components: one corresponding to the host galaxy and one to a typical QSO (Fig. \ref{fig.modelfit}). As QSOs are typically 
found to be hosted by massive elliptical galaxies \citep{mcleod1995, aretxaga1998, dunlop2003}, we adopted a 5 Gyr early-type 
galaxy template, generated with the GRASIL code \citep{silva1998}. To reproduce the QSO we adopted a composite spectrum of a 
type 1 QSO with the highest IR/optical ratio \citep[from][]{polletta2007}. In the model the QSO SED component has intrinsic 
absorption, i.e. nuclear extinction, parameterised by a rest-frame equivalent $A_V$ and represented by the Small Magellanic 
Cloud (SMC) extinction curve from \citet{pei1992}, whilst both the QSO and galaxy SEDs have fixed Galactic absorption 
($N_H^{gal}=2\times10^{20}\ \rm cm^{-2}$), using the Milky Way extinction curve \citep{pei1992}. The other free parameter in 
the model is the \gf\ ($g$), which we define as the ratio between the host galaxy and the intrinsic QSO fluxes in the 
rest-frame wavelength range $\lambda_1=4160\ \AA<\lambda<\lambda_2=4210\ \AA$, chosen because it avoids most of the galaxy 
absorption lines and the quasar emission lines \citep{vandenberk2006}, i.e.:
\begin{displaymath}
g=\frac{\int_{\lambda_1}^{\lambda_2} f_{\lambda,\ gal}\ d\lambda}{\int_{\lambda_1}^{\lambda_2} f_{\lambda,\ \rm QSO}\
d\lambda}.
\end{displaymath}

\begin{figure}
\centerline{
\includegraphics[scale=0.3,angle=-90]{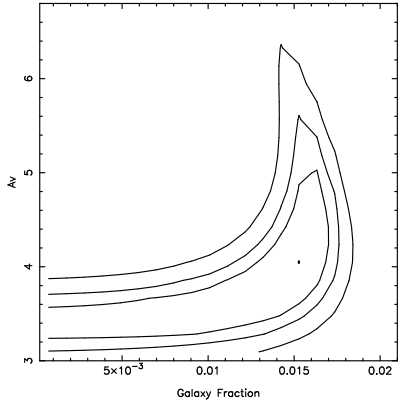}}
\caption{$\chi^2$ contours at 68, 90 and 99\% confidence obtained from modelling of the IR spectrum for a grid of \gf\ 
and A$_V$ values, as defined in the text.}
\label{fig.model}
\end{figure}

We used this SED model to fit the IR continuum (Fig.~\ref{fig.irspec}) in the wavelength range 14500-23000 \AA. As we  aimed to 
reproduce only the spectral continuum, we excluded the emission line from the fitting; we excluded also the wavelengths below 
14500 \AA\ because of the lower S/N in this part of the spectrum and to avoid the wavelengths of the expected [O{\sc III}] and 
the H$\beta$ emission lines. After binning the spectrum, we computed $\chi^2$ values for a grid of possible nuclear extinction 
and \gf\ ($g$) parameters and we determined confidence intervals for the two free parameters using the usual $\Delta\chi^2$ 
prescription (Fig. \ref{fig.model}). Our modelling provides strong constraints on the rest-frame equivalent $A_V=3.2-5.0$ (90\% 
confidence), while the galaxy contribution is not strongly constrained. The best-fit of the near IR spectrum and the residuals 
of the fitting are shown in Fig. \ref{fig.modelfit}. We can also use the model fit results to constrain the expected $r'$, $i'$ 
and $J-$band magnitudes. Our best-fit model predicts $r'\approx 27.7$, $i'\approx 26.3$ and $J\approx 22.6$, values consistent 
with the upper limits discussed in \S\ \ref{opt}. We note that the general properties of \src\ are rather similar to the
object presented by \citet{severgnini2006}, although our source has higher extinction and has not yet been detected in the 
optical band. Our constraints on the host galaxy mass are comparable to, or somewhat lower than, that inferred by 
\citet{severgnini2006} for their object.

Our modelling cannot of course be considered as definitive, as it is limited by the restricted wavelength coverage of our data, 
by the fact that we cannot be sure of the intrinsic SEDs for the galaxy and AGN components and by the assumption that just two 
components will adequately represent the data. We can however examine how our results depends on the choice of the SED 
components in the spectral range we are investigating, by using different QSO templates (from \citealt{polletta2007}) in the 
fitting process. For this range of templates we find variations of $\lesssim10$\% in the derived rest-frame equivalent $A_V$. As
in all the models we tested the contribution of the galaxy appears to be only few percent of the total emission there is clearly
little sensitivity to the precise shape of the galaxy SED adopted.

With the model parameters obtained from our analysis, the extinction for the QSO component in the $J-$band (close to H$\beta$ 
and [O{\sc III}]) corresponds to $\approx5.3$ mag and $\approx2.4$ mag in the $K-$band, in the vicinity of the redshifted 
H$\alpha$ line. Correcting the observed H$\alpha$ flux for this extinction ($A_K$) gives a value 
$f_{\rm H\alpha}\approx4.6\times10^{-15}$ \cgs, which is lower than the flux expected from the $L_{\rm X}$-$L_{\rm H\alpha}$ 
correlation of \citealt{panessa2006} (see Table \ref{table2}), but well within the scatter of the correlation. The much higher
extinction ($A_J\approx5.3$) at the H$\beta$ line is fully consistent with its non-detection.
\begin{table}
\begin{tiny}
\caption{Observed and predicted line fluxes of \src.}
\label{table2}
\begin{tabular}{c c c c c}
\hline\hline
\rule[-1.5mm]{0pt}{4ex}Line& Observed Flux$^a$& Predicted Flux$^b$ & Corrected Flux & $A_\lambda^c$ \\
\rule[-1.5mm]{0pt}{3ex} & [\cgs] & [\cgs] &  [\cgs] & [mag] \\
\hline
\rule[-1.7mm]{0pt}{4ex} H$\alpha$    & $(5\pm0.2)\times10^{-16}$ & $1.7\times10^{-14}$ & $4.6\times10^{-15}$ & 2.4\\
\rule[-1.7mm]{0pt}{4ex} H$\beta$     & $<1\times10^{-16}$        & $4.8\times10^{-15}$ &      $-$            & 5.3\\
\rule[-1.7mm]{0pt}{4ex} [O{\sc III}] & $<3\times10^{-17}$        & $2.9\times10^{-15}$ &      $-$            & $-$\\
\hline
\end{tabular}

$^a$ Uncorrected for extinction, measured from the IR spectrum.\\
$^b$ From the correlations of \citet{panessa2006}.\\
$^c$ Extinction estimated from the model.
\end{tiny}
\end{table}

The lack of detection of the [O{\sc III}] line is not explained in the model, as this line is expected to originate in the 
Narrow-Line Region (NLR) which should not be affected by the large extinction of the nuclear region. However, the 
weakness or disappearance of the [O{\sc III}] emission line has already been reported for high-luminosity AGN 
\citep[see][]{yuan2003,netzer2004,sulentic2004}. A possible explanation may be related to the fact that the simple scaling law:
$R_{NLR}\propto L_{ion}^{1/2}$ (where $R_{NLR}$ is the radius of the Narrow Line Region and $L_{ion}$ is the ionising source 
luminosity, \citealt{netzer2004}), which may explain the correlation between the X-ray and the [O{\sc III}] luminosities at 
lower values, must break down when the NLR radius becomes comparable with the size of the galaxy. 

Another possible explanation may be the presence of obscuring dust {\it outside} the torus, at larger distances from the 
nucleus. This dust could reside in the NLR \citep{polletta2008} or in the host galaxy \citep{rigby2006,alejo2006,brand2007} and 
could thus absorb the emission coming from the Narrow Line Region. This interpretation is equally consistent with the observed 
red continuum and the extinction of the broad lines whilst also providing an explanation for the absence of any narrow lines in 
the IR spectrum. However, as it is not possible to constrain the location of the obscuring dust with the present data, neither 
interpretation can be rejected.

Finally we note that the values obtained for the optical extinction from our modelling are significantly lower than naively 
expected from the measured X-ray column density. This can most easily be parameterised by a dust-to-gas ratio which is about
$20-25$ times lower than  the standard Galactic value ($A_V=N_H/(1.8\times10^{21})$, e.g. \citealt{predehl1995}). 
Such low ratios have been reported in  several previous studies of AGN \citep{maccacaro1982, maiolino2001b, akiyama2002,
willott2004}. In this comparison, as is the case in the other papers cited, we are assuming the simplest possible 
geometry where the dust and gas are co-spatial and lie in a uniform foreground screen\footnote{This assumption may lead to 
an underestimation of the effective extinction of the source \citep[e.g.][]{mackenty2000}. More realistic dust and gas 
distributions can of course lead to different conclusions about the dust-to-gas ratio; however, as it is not possible to 
constrain the gas and dust distribution with the present data, considering more sophisticated models is beyond the purpose of our
analysis.}. One possible explanation for the high ratio is a different dust grain size distribution dominated by large grains, 
whose formation is naturally expected in the high density environments, like those characterising the circumnuclear region of 
AGNs. This dust grain distribution makes the extinction curve flatter than the Galactic one and yields a higher 
$N_H/A_V$ ratio \citep{maiolino2001a,maiolino2002}. This would not be a valid explanation if the absorption is 
on kpc-scales as discussed above.

\section{Summary and conclusions}
\label{conc}
On the basis of its X-ray and optical properties, \src\ has the highest recorded \fxo\ of any extragalactic X-ray source. IR 
spectroscopy confirms this object is an AGN at $z=1.87$. In X-rays it is a very luminous object with large absorption, making it
a type 2 QSO. Its optical/NIR properties show a strongly reddened continuum, but stellar appearance in the $K$ band and broad 
H$\alpha$, more consistent with a type 1 classification (type 1.9 in the classification used for low redshift AGN). We have 
shown that these apparently discrepant properties can be reconciled with a model consisting of a heavily reddened AGN nucleus 
(rest-frame equivalent $A_V=3.2-5.0$) and fainter host galaxy. Our results provide tight constrains on the X-ray column density 
and optical/IR extinction which demand a dust-to-gas ratio for this object $\sim25$ times lower 
than the standard Galactic value. The non-detection of a narrow [O{\sc III}] emission line at anything like the expected flux 
in this object may be related to its very high luminosity.

Alternatively \src\ might be a member of a population of AGN with absorption on kpc-scales, sometimes described as 
``host-obscured" AGN \citep[e.g.][]{brand2007}, which may be an important ingredient in resolving the discrepancy between the 
predicted and observed ratios of type I and type II AGN, which exists in some models of AGN obscuration 
\citep[e.g.][]{alejo2006,brand2007}.

We have shown that \fxo\ selection using the large samples afforded by the 2XMM catalogue provides an effective way of 
discovering extreme objects like \src, which are a rare but important part of the obscured AGN population. Our analysis has 
demonstrated that the extreme properties of this object are a natural consequence of its very high luminosity and large 
obscuration. It is interesting to note that a high luminosity object of this type with only a factor of a few higher 
absorption would appear as an entirely ``normal" galaxy, in that it would have no detectable broad lines in its spectrum and 
presumably no narrow lines either, given the apparent strong suppression of the NLR emission lines evident at these high 
luminosities.

\begin{acknowledgements}
The data presented here include those taken using ALFOSC, which is owned by the Instituto de Astrof\'\i sica de Andaluc\'\i a 
(IAA) and operated at the Nordic Optical Telescope under agreement between IAA and the NBIfAFG of the Astronomical Observatory 
of Copenhagen. The analysis in this paper are based in part on data collected at Subaru Telescope, which is operated by the 
National Astronomical Observatory of Japan. The United Kingdom Infrared Telescope is operated by the Joint Astronomy Centre on 
behalf of the Science and Technology Facilities Council of the U.K. We gratefully acknowledge the SPARTAN support under the 
contract MEST-CT-2004-7512. FJC acknowledges financial support by the Spanish Ministerio de Educaci\'on y Ciencia under project 
ESP2006-13608-C02-01. SM acknowledges direct support from the U.K. STFC research Council. We thank the referee for the 
constructive comments, which helped to improve the paper.

\end{acknowledgements}

\bibliographystyle{aa}
\bibliography{../allpapers}

\end{document}